\documentclass[prb]{revtex4}
\usepackage{graphicx}
\usepackage{psfrag}

\psfrag{OmegaV}{$\Delta\Omega/\rho_0 a^3 \sqrt{N} kT$}
\psfrag{radius2}{$r_0^2/Na^2$}

\begin{document}

\title{Theory of copolymer micellization}

\author{ Daniel Duque}

\affiliation{%
Department of Physics\\
Simon Fraser University, Burnaby BC, Canada}

\date{\today}

\begin{abstract}%
We consider the micellization of
block copolymers in solution,
employing self consistent field theory with
an additional constraint that permits the examination
of intermediate structures.
From the information for an isolated micelle
(structure, binding energy, free energy)
we describe how the global thermodynamics of these
systems can be obtained,
which can be used to build a realistic phase diagram;
the role of translational entropy must be addressed in this regard.
\end{abstract}

\maketitle

\section{Introduction}

Amphiphilic molecules, presenting
simultaneous solvophobic and solvophilic interactions,
form many interesting structures in solution in
which the solvophobic residues are screened from the solvent, such
as bilayer structures and micelles.
The later aggregate is remarkable, since,
as they begin to
proliferate around what is
known as the critical micelle concentration (CMC),
its size does not become macroscopic, but is limited by the
particular molecular features.
It is the main goal
of this paper to provide a unifying view of micelle formation
for the case of copolymers in solution \cite{hamley},
in the regime where the micelle concentration is low, so that
each micelle can be regarded as isolated;
the translational entropy of the aggregates
can play an important role in this regime, as we will discuss.
(A dense fluid will of course
be more complicated, since the interactions between micelles
must be taken into account~\cite{dormidontova,TDC}.)

We will use  a simple version of
self consistent field theory (SCFT) \cite{MatsenSCFT,schmid}.
The use of self-consistent theories for
copolymer micelles has a long history, beginning with
the work of Yuan et al. in 1992 \cite{yuan}.
Our approach differs from previous ones in that we are
not primarily interested in the calculation of properties of
isolated micelles, but rather in how to include this
information in a thermodynamic description of the micellar
fluid. Even though the importance of the excess free energy of the micelles
in defining a proper CMC and distinguishing between
competing structures
has been recognized since the first works~\cite{shull1993},
we feel the connection between it and the concentration
of micelles, which can lead to differences
between possible definitions of the CMC, has not been
explored in detail.
Additionally, we also employ the method of
Refs.~\onlinecite{matsen3,mueller} in order
to explore intermediate structures, that lie between
equilibrium ones.

This method can be classified with previous approaches
to obtain the general thermodynamics from
the properties of a single aggregate.
There are several studies that focus on
planar and curved bilayer\cite{CST} or monolayer \cite{matsen3}
membranes, obtaining quantities like
surface tensions and bending rigidities;
micelles can be considered as limiting cases of
these structures, as is discussed in some cases\cite{SCMT,oxtoby}.
There are studies focused
mainly on micellar structure and thermodynamics \cite{DT,linse},
but few consider a global treatment.
(Refs.~\onlinecite{homopolymer,noolandi} come quite close,
but are mainly concerned with periodic dense phases.)

In section \ref{isolated} we calculate properties of individual
micelles; we discuss how to build the global thermodynamics of
the system in section \ref{thermo}.

\section{Isolated micelles}
\label{isolated}

We have decided to focus on
a copolymer--homopolymer mixture \cite{homopolymer},
using the standard Gaussian model for polymers
treated within SCFT,
perhaps the simplest molecular theory
that produces realistic results for micellization (just as
it is, arguably, the simplest one that produces realistic
complex mesophases).
This way, we can avoid the complications associated with
similar models that are more refined and specific
\cite{noolandi,shull2002},
while keeping a molecular theory that is closer to experimental
systems than other models based on hard sphere fluids
\cite{DT,SCMT,CST,oxtoby}.

We consider an incompressible
mixture of copolymers and homopolymer
with volume $V$. The copolymers have $N$
statistical units of volume $\rho_0^{-1}$ and statistical length $a$;
these are of type A from one end to monomer $f N$ and type B from there on.
The homopolymer is made of monomer A and has $\alpha N$ units of
the same volume and length.
In a system of volume $V$ there will be an overall homopolymer
volume fraction $\phi$, which will be the spatial average of
a homopolymer distribution $\phi(r)$.
By incompressibility, the
corresponding copolymer volume fraction will be $1-\phi$, and
its distribution, $1-\phi(r)$.
Notice the overall copolymer
concentration is $\rho_c=\rho_0 (1-\phi)/N$, and the homopolymer
one, $\rho_h=\rho_0 \phi/(\alpha N)$.

We will consider a grand free energy
\begin{eqnarray}
\label{grand}
\frac{ N \Omega }{k T \rho_0 V} &=&
-(z q_h + q_c)+
\chi N  \int dr \phi_A (r)\phi_B(r)+
\int dr \left[ \phi_A(r) w_A(r) + \phi_B(r) w_B(r) \right] \\
& &+ \int dr \left[
\xi(r) (\phi_A (r)+ \phi_B(r)-1)+
\psi (\phi_A (r) - \phi_B(r) )\delta(r-r_0)  \right].
\end{eqnarray}
The first term in the right side is the configurational entropy,
$q_h$ and $q_c$ being the partition functions of a homopolymer
and copolymer in the corresponding fields and $z$
the fugacity of the homopolymer 
(by incompressibility, that of the copolymers can be taken to be $1$,
shifting the scale of chemical potentials).
The second term is the interaction energy; the
third contains the coupling with the fields $w_A$ and $w_B$; the fourth,
a Lagrange multiplier, $\xi(r)$, to enforce incompressibility.
The last term introduces an additional variable $\psi$ which lets us sweep a range
of metastable structures by ``pinning'' the profiles to a
certain value at some point in space $r=r_0$.
(This goes beyond standard SCFT,
as was used for this model in Ref.~\onlinecite{homopolymer} even though this
technique was used by the same author in Ref.~\onlinecite{matsen3}
for a similar system;
see also Ref. \onlinecite{mueller} for an application to
related structures, spherical nucleation bubbles, and
Ref. \onlinecite{kirill}, in
which such a field is helpful to study bilayer fusion;
in Ref.~\onlinecite{gersappe} hards wall are used in order
to explore different micellar sizes.)

The extremization of the grand free energy with
respect to all the volume fraction profiles and fields
leads to the self-consistent equations that have to be
solved, with the copolymer and homopolymer profiles related
to the fields through Green functions, $q_c(r,s)$, $q_h(r,s)$,
which satisfy the standard diffusion equations \cite{homopolymer}
and also provide the partition functions
$V q_c =\int dr q_c(1)$, $V q_h =\int dr q_h(\alpha)$.
This kind of theory is well know to produce rich phase diagrams,
with many periodic structures. But it is
also possible to obtain certain structures that are localized.
In these, the volume fraction profiles of the different components
tend to their bulk values away from the spatial point where
the structure is located. The bulk values are those of
the corresponding disordered phase:
$\phi(r)\rightarrow\phi^b$,
$\phi_A(r)\rightarrow\phi^b+f(1-\phi^b)$, and
$\phi_B(r)\rightarrow (1-f)(1-\phi^b)$.
These localized structures correspond to isolated
bilayers and micelles.

The theory is simple to implement in planar, cylindrical and spherical geometry,
but we will focus on the later in this article.
The diffusion equations have been solved in
real space using Cranck-Nicholson's method,
which is much simpler conceptually (even if
not so efficient computationally) than Fourier methods.
(This procedure is also useful for problems in which the geometry is not
known \cite{fgd} or which are not very symmetric \cite{kirill}.)
With the help of the additional $\psi$ field, we study spherical micelles,
as well as larger, metastable structures, corresponding to spherical bilayer vesicles.
Notice this choice of $r_0$ as a ``reaction coordinate''
has some problems. Bilayer structures have two points at which $\phi_A(r)=\phi_B(r)$;
for spherical and cylindrical geometries,
it is preferable to assign $r_0$ to the outer one (the
one for which $r$ is larger),
since the inner one will be seen to disappear as the profiles
become micellar.
More importantly, for some
values of $r_0$, typically at regions of
transition between different morphologies,
no solution to the SCFT
is found, a fact we will discuss below.

We employ the iterative
method described in Ref.~\onlinecite{fgd}, modified to include
the $\psi$ field, to solve the self-consistent equations.
It is found that $200$ space points per $a\sqrt{N}$ and
$4000$ points along the chain
are enough to provide results that
cannot be distinguished from previously published
results. A system size of $5 a\sqrt{N}$ is sufficient
to make finite size effects negligible.

\begin{figure}
\begin{center}
\includegraphics[width=10cm]{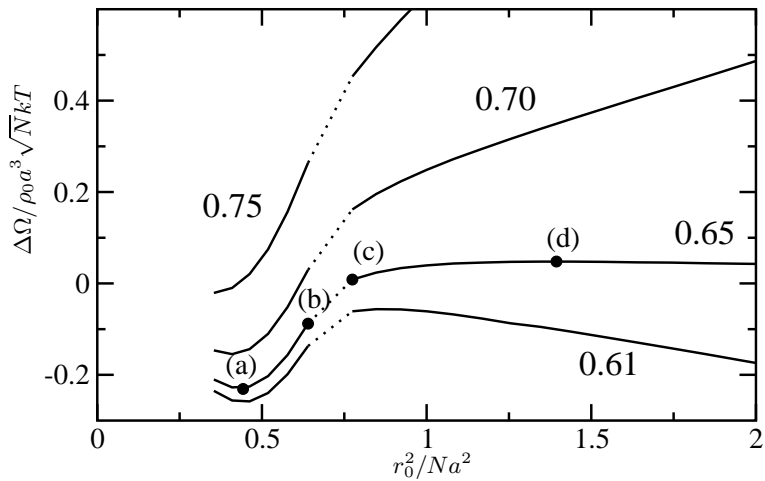}
\caption{%
Excess grand free energy as
a function of $r_0^2$ for values of
$\phi^b= 0.61, 0.65, 0.70, 0.75$.
The dotted lines span the range for which
no solutions are found.
The points correspond to the selected values
that are used in Figure~2.}
\end{center}
\end{figure}

We will choose this set of parameters:
$\chi N=12$, $f=0.6$, $\alpha=1$;
in Ref.~\onlinecite{homopolymer} we see this system shows,
at high homopolymer concentrations, an ``unbinding''
of a micellar bcc crystal into isolated spherical micelles.
In Fig.~1 we plot $\Delta\Omega$, the
excess grand free energy, as a function
of the $r_0^2$ for several values of $\phi^b$, for spherical symmetry.
We can see that for larger values of $\phi^b$ (i.e., low
values of bulk copolymer), there are no nontrivial stable or metastable
solutions. Above a certain value, there appears a minimum, which
corresponds to a spherical micelle (this is the value associated
with the CMC in some works).
The $\Delta\Omega$ is still positive, but becomes negative at
lower $\phi_b$, a value that would be close to the CMC. Then, at
even lower values, the asymptotic straight lines change slope.
This linear behavior, for large aggregates, is characteristic of
spherical vesicles, and, as described below,
the corresponding profiles are indeed vesicular.
The free energy as a function of the area $A$ will tend to
a line \cite{SCMT,matsen3}:
\(
\Delta\Omega\rightarrow \gamma A + 8\pi (\kappa+\kappa_G),
\)
with slope $\gamma$, the surface tension;
$\kappa$ is the bending rigidity and $\kappa_G$ is the
Gaussian bending rigidity. That is,
the change in slope corresponds to a change in the surface
tension of the membrane; indeed, the condition for the stability of
a membrane is $\gamma=0$. There usually is some leeway regarding
the definition of the area $A$ (i.e., we could propose
$A=4\pi r_0^2$, hence our choice of $x$ axis),
but when $\gamma=0$ there is none.
This method can be used to obtain the rigidities, since
the cylindrical geometry can separately provide $\kappa$ \cite{matsen3,CST}.
In this particular case, though, the lamellar transition
is preempted, since at this point the $\Delta\Omega$ of the micelles
is quite negative, as is clearly seen in the curve for
$\phi_b=0.65$ in Fig.~1.
This is just what would be expected, since our CMC is just
the ``cubic phase unbinding'' of Ref.~\onlinecite{matsen3}, i.e.,
the stable periodic phase corresponding to our choice
of parameters is a cubic phase, not a lamellar one.

In Fig.~2 we present some typical profiles for the
case $\phi^b=0.65$, at the points in Fig.~1.
Fig.~2(a) corresponds
to the stable micellar structure, with
a core composed of B copolymer and a
corona of A copolymer.
As $r_0$ is increased the micelle grows,
Fig.~2(b) shows the most ``swollen'' micelle
we are able to obtain, still with the same
architecture.
After a region of $r_0$
for which no stable profiles are found,
the next stable structures show a core
composed of the A region of the copolymer and
a little solvent, Fig.~2(c).
This sort of ``proto-vesicle'' is a micelle with A-B-A structure, instead
of the previous B-A one.
For larger values of $r_0$ homopolymer progressively
fills the core (which is now
of an A nature, and thus compatible with it),
while the copolymer profiles tend to the planar A-B-A bilayer structure as
in Fig.~2(d), which corresponds to a local maximum
in the excess free energy.

\begin{figure}
\begin{center}
\includegraphics[width=10cm]{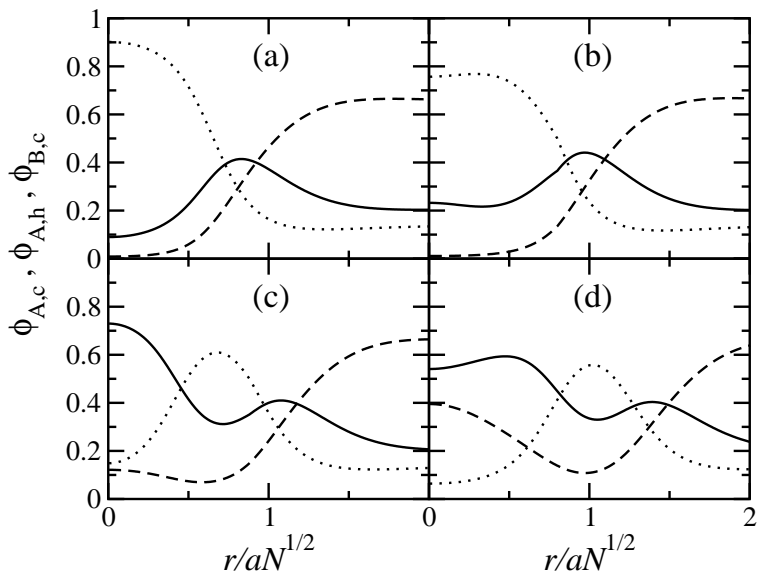}
\caption{%
Selected volume fraction
profiles for $\phi_b=65$, points
highlighted in Fig.~1:
(a): $r_0^2=0.44 Na^2$ (equilibrium);
(b): $r_0^2=0.64 Na^2$;
(c): $r_0^2=0.77 Na^2$;
(d): $r_0^2=1.40 na^2$.
Solid lines:
copolymer A; dashed: homopolymer A;
dotted: B.}
\end{center}
\end{figure}

We would like to point out it is possible to connect
this results with an  approximation proposed in Ref.~\onlinecite{oxtoby},
to obtain the whole set of curves for unstable
structures from information about critical ones (i.e., at local
minima or maxima). Our results show that this
approximation works quite well for large radii, but breaks down
at  small ones.
In fact, an examination of the structures shows that the collapse
of micellar structures at lower values of $\phi^b$ does not take
place because they become unstable with respect to planar membranes
(as is the case in Ref.~\onlinecite{oxtoby}), but because the whole system
begins to develop long range oscillations --- that is, because we
reach the spinodal.

\section{Global thermodynamics}
\label{thermo}

We have seen one of the most important micellar quantities is
its excess (grand) free energy, $\Delta\Omega$;
another one is the excess solvent, $\Delta\phi$, a negative quantity
(the excess of copolymer is $-\Delta\phi$, positive).
At low micelle concentrations,
the components are either in the bulk or in micelles, so the
total concentration of copolymer
\(
\rho^t_c=\rho^b_c+\rho_m \Delta N_c,
\)
where $\rho^b_c$ is the bulk concentration,
$\rho^b_c=\rho_0(1-\phi^b)/N$, $\Delta N_c$ is
the excess of copolymer molecules in
the micelle (proportional to $|\Delta\phi|$, as we discuss below) and
$\rho_m$ is the concentration of micelles.
This quantity is expected to satisfy \cite{DT}
\begin{equation}
\label{rhom}
\rho_m=\frac{1}{v_m}\exp(-\Delta\Omega/kT),
\end{equation}
which is dominated by the condition $\Delta\Omega=0$,
but the value of the volume $v_m$, which is related
to the translational entropy of the micelles, can make a
difference, leading to different choices for the CMC.
(Another
important discrepancy is that many authors choose
to relate the CMC with the point at which the micelles become
stable (in general, metastable) \cite{oxtoby,linse,noolandi,girardi,DD}.)
We propose, based on our previous work~\cite{DT}:
\(
 v_m= (\rho_0/N) \exp[(\Delta E-\Delta F)/(k T \Delta N)],
\)
where $\Delta E$ is the excess interaction energy, (second term in Eq.~\ref{grand}
minus the bulk energy), $\Delta F$ is the excess Helmholtz free energy
($\Delta F=\Delta \Omega+\mu (\Delta\phi)/\alpha$), 
and $\Delta N$ is the total number of molecules in a micelle.

We briefly discuss two problems that can be encountered when
calculating $\rho_m$.
First, all the densities are
given in units of $\rho_0/N$, and all free energy densities in
units of $k T \rho_0/N$; on the other hand,
a well know feature of the Gaussian model is that
the spatial
variation of the profiles is set by the combination $a\sqrt{N}$.
This means that,
in order to obtain $\Delta\Omega/kT$, we need the ratio
between the two measures of volume,
\(
\lambda \equiv (a\sqrt{N})^3/(N/\rho_0)=\rho_0 a^3 \sqrt{N};
\)
i.e., how many polymers would there be in a volume 
$(a\sqrt{N})^3$. This is basically the parameter that describes
the degree of concentration of a polymer solution, and by
definition, $\lambda\gg 1$ for a melt, but this value must be
provided for each particular system.
(Remarkably, our expression for $v_m$ does not depend on
$\lambda$.)

Second, the calculation of $\Delta N$ is not obvious.
There is always some arbitrariness in defining this
magnitude; for copolymers, it is natural to propose
$\Delta N_c=\lambda|\Delta\phi|$.
But the micelles also include a certain number of solvent molecules (specially in the corona region).
Looking at the
profiles in Fig.~2, this would be the integral of the homopolymer profiles,
dashed lines (together, of course, with a factor of $4 \pi r^2$),
but restricted to the micellar region, since this integral diverges, unlike
the one for copolymers.
We propose to
use a weight $w(|\phi(r)-\phi^b|)$, where $w(x)$ is a function
with a limit $w(x)\rightarrow x$ for small $x$ but 
$w(x)\rightarrow 1$ for large $x$ 
to ensure the integration is properly restricted.

Finally, we would like to comment that we have compared our calculation
with a previous lattice model~\cite{linse}.
The authors use a form of SCFT, and Eq.~(\ref{rhom}) (including a proposal
for $v_m$).
The systems is slightly different, since they consider a simple
solvent, not a homopolymer one, but
the proper theory is very similar to the one written above:
all the equations remain the same, 
except that the diffusion
equations become simple Boltzmann distributions,
since a solvent molecule has no internal conformations:
$\phi_{A,s}(r)=\alpha z q_s(r)$ and
$q_s(r)=\exp[-\alpha w_A(r)]$.
The appropriate choice of our parameters to mimic the lattice model
is: $f=1/2$, $\alpha=1/100$, $\chi N=124$, and
$\lambda=10$. Our results deviate markedly from
this Reference: the location of the CMC is quite close, but
in our case the volume fraction of micelles is high as soon as they
become stable, so the assumption of isolated
micelles breaks down, and one should consider instead the dense, periodic
phase (very likely, a cubic phase) as the proper equilibrium structure.
This difference is likely due to the
lattice description of  chain conformations in  Ref.~\onlinecite{linse},
which can lead to an underestimation of the corresponding entropy
and hence to a prediction of a much smaller micellar concentration.
In addition, we have checked that
spherical micelles are indeed the stable structure,
by comparing with cylindrical ones and planar bilayers.

\section{Acknowledgements}

We warmly thank Dr. Michael Schick for
making this work possible. 
We also thank Dr. Pedro Tarazona for his help
during a stay in which this work was initiated and greatly advanced,
and Dr. Kirill Katsov for enlightening discussions.
Financial support was provided by
the Ministry of Education, Culture and Sports of Spain, with a
MECD2000 grant.

\bibliography{mic}

\end{document}